\documentclass[
reprint,
superscriptaddress,
showpacs,
preprintnumbers,
amsmath,
amssymb,
aps, aip, prl
floatfix, 
] {revtex4-2}

\usepackage{graphicx}  
\usepackage{dcolumn} 
\usepackage{bm}        
\usepackage{float}
\usepackage{siunitx}
\usepackage{xcolor}
\usepackage{comment}
\usepackage{bm}
\usepackage{amsthm}
\usepackage{xcolor} 
\usepackage{soul}   
\usepackage{upgreek}

\begin{document}

\title{Experimental Scaling of Diffraction Efficiency in Laser-Induced Plasma Gratings}

\newcommand{\equalcontrib}{These authors contributed equally to this work.}

\author{M. M. Wang }
\email{michelle.wang@princeton.edu}
\affiliation{Princeton University, Princeton, NJ, 08544 USA}

\author{V. M. Perez-Ramirez}
\thanks{M. M. Wang and V. M. Perez-Ramirez contributed equally to this work.}
\affiliation{Stanford University, Stanford, CA, 94305 USA}

\author{N. M. Fasano}
\affiliation{Princeton University, Princeton, NJ, 08544 USA}

\author{K. Ou}
\affiliation{Stanford University, Stanford, CA, 94305 USA}

\author{S. Cao}
\affiliation{Stanford University, Stanford, CA, 94305 USA}

\author{V. Dewan}
\affiliation{Princeton University, Princeton, NJ, 08544 USA}

\author{A. M. Giakas}
\affiliation{Princeton University, Princeton, NJ, 08544 USA}

\author{A. Morozov}
\affiliation{Princeton University, Princeton, NJ, 08544 USA}

\author{P. Michel}
\affiliation{Lawrence Livermore National Laboratory, Livermore, CA, 94551 USA}

\author{M. R. Edwards}
\email{mredwards@stanford.edu}
\affiliation{Stanford University, Stanford, CA, 94305 USA}

\author{J. M. Mikhailova}
\email{j.mikhailova@princeton.edu}
\affiliation{Princeton University, Princeton, NJ, 08544 USA}

\begin{abstract}

We demonstrate efficient diffraction of intense ultrashort laser pulses using optical-field-ionization-induced plasma-neutral gratings formed by spatially structured ionization of a neutral molecular gas in the interference field of two femtosecond pump pulses. The transient refractive index modulation of the plasma structure persists for  at least 10 picoseconds and is used to diffract intense femtosecond signal pulses into the 1st order of diffraction with an average efficiency of up to 35$\%$. Plasma gratings are shown to provide stable diffraction at signal laser intensities greater than $ 10^{14}\text{ W/cm}^2$, exceeding the damage thresholds of conventional solid-state optics by more than two orders of magnitude, continuously for hours at a 10-Hz repetition rate. The experimental diffraction efficiency scales with the grating aperture allowing for a larger millimeter-scale plasma optic, increases with the pump energy and electron density, and reaches a maximum at a specific grating length in agreement with the coupled-mode theory for periodic media. These results demonstrate the scalability, tunability, and high damage threshold of transmissive plasma-based photonic structures, opening new prospects for controlling multi-petawatt laser beams.

\end{abstract}

\maketitle

\section{Introduction}

Ultra-intense lasers reaching multi-petawatt peak powers are driving laser-based particle acceleration, high-energy photon sources, inertial fusion energy, and new regimes of nonlinear quantum electrodynamics \cite{Pirozhkov2017, Wang2017ALEPH, Guo2018, Gales2018, Bromage2019, Yoon2021, Veisz2025, maksimchuk2025zeus, AbuShawareb2024, Los2026, DiPiazza2012}. However, further increases in peak laser power and intensity are constrained by laser-induced damage of solid-state optical components, which is difficult to mitigate due to the inherently stochastic spatial and temporal structure of high-power laser beams \cite{Zheltikov2024_PRA, Zheltikov2024_PhysLettA}. 

 In contrast to conventional solid-state transmissive optics, where discrete quantum resonances of bound charges dominate the optical response and functionality, and the optical damage by pulsed lasers at high laser fluences is primarily due to ionization of the optical material, transient plasma-based optics make use of the collective response of already ionized free electrons created by laser-induced ionization in gases, liquids, or solids. Plasma structures can withstand higher intensities without degradation, allowing plasma-based optical components to be reduced in size from meters to millimeters for petawatt beams. The optical response of plasma electrons is intrinsically ultrafast and broadband \cite{Mikhailova2025}, as it is largely determined by the plasma frequency, which depends on the free electron density, and is therefore not restricted to specific wavelengths defined by quantum resonances \cite{edwards2021laser}. By tailoring the electron density, plasma-based optical components can be engineered for high-power lasers across a wide spectral range from near-infrared petawatt \cite{Pirozhkov2017, Wang2017ALEPH, Guo2018, Gales2018, Bromage2019, Yoon2021, maksimchuk2025zeus} to mid-infrared multi-terawatt systems \cite{polyanskiy2020demonstration}, ultra-broadband multicolor waveforms \cite{Edwards2020_Two_color, Veisz2025}, and extreme-ultraviolet and x-ray sources \cite{Edwards2017,branlard2012european,Kramer2023, fasano2023harmonic}. In addition, the plasma-optic medium renews with each laser pulse, therefore damage does not accumulate even at high repetition rates.
 
 Recent advances in plasma optics include ultrafast optical switches for temporal contrast cleaning \cite{Mikhailova2011,edwards2024greater}, relativistic plasma mirrors for high-energy photon sources \cite{thaury2007plasma,Heissler2012,Mikhailova2012,edwards2014enhanced,edwards2016waveform,edwards2020x,fasano2024cascaded,fasano2024plasma,Haessler2022,Jahn2019,Chopineau2022,Ouill2024,Fasano2023,Edwards2020, edwards2016multipass} and for on-target diagnostics of intense ultrashort pulses \cite{Leshchenko2019}, plasma waveplates \cite{michel2014dynamic,qu2017plasma}, optical-field-ionization waveguides for laser wakefield acceleration \cite{Morozov2018,Feder2020, lemos2018guiding,Picksley2024,Rankin1991}, compressor gratings for chirped-pulse amplification \cite{wu2005manipulating,edwards2022plasma,lehmann2024plasma,wang2025experimental}, and holographic diffractive lenses \cite{palastro2015plasma,edwards2022holographic}, highlighting plasma optics as a rapidly developing area in high-power laser science and technology.

The formation of transient laser-induced plasma structures can be governed by either ponderomotive density modulation in ionized plasmas \cite{sheng2003plasma,lehmann2016transient, edwards2022plasma} or spatially varying ionization in a neutral gas \cite{durand2012dynamics,shi2011generation,zhang2021ionization,liu2011two,suntsov2009femtosecond, edwards2024greater}. Here we focus on the latter mechanism, where spatially periodic optical field ionization creates a refractive index modulation between plasma and neutral gas regions. In our experiments, thick plasma–neutral gratings operate in the Bragg regime, where most of the incident energy is diffracted into the first order at the Bragg angle, defined by the criterion $\rho \gg 1$, with $\rho=\lambda^2/(\Lambda^2 n_0 n_1)$, where $\lambda$ is the incident wavelength, $\Lambda$ is the grating period, $n_0$ is the mean refractive index, and $n_1$ is the amplitude of the fundamental refractive index modulation for a sinusoidal grating \cite{moharam1978criterion}.

The highest reported diffraction efficiency from femtosecond optical field ionization gratings in a neutral gas was $\sim19\%$, achieved with 100 fs ultraviolet (267 nm) pulses producing peak electron densities of $\sim10^{19}$ cm$^{-3}$ in air \cite{shi2011generation}. Plasma–neutral gratings have also been generated through collisional ionization with longer picosecond pulses, yielding average efficiencies up to $36\%$ using 1064 nm, 120 ps, $<250$ mJ pump pulses that produced plasma densities approaching $8\times10^{18}$ cm$^{-3}$ in CO$_2$ gas to diffract a 3.9~$\upmu$m signal beam \cite{edwards2023control}. Higher diffraction efficiency is expected for longer signal wavelengths due to the inverse-square scaling of the critical plasma density with the signal beam wavelength. The plasma refractive index is $n_{\mathrm{plasma}}=\sqrt{1-N_e/N_c}$, where $N_e$ is the electron density and $N_c=\epsilon_0 m_e \omega_0^2/e^2 \propto 1/\lambda_0^2$ is the critical plasma density. The resulting increase in refractive index contrast between plasma and neutral regions, combined with an optimal grating length, can enhance the diffraction efficiency for longer signal wavelengths.

Here we demonstrate high diffraction efficiency of field-ionized plasma–neutral gratings over a range of grating parameters. Using near-IR (800 nm), 29 fs pump pulses with mJ-level energies, we generate ionization gratings with peak electron densities of only $\sim 4\times10^{17}$ cm$^{-3}$, which is more than an order of magnitude lower than in previous studies \cite{shi2011generation, edwards2023control}, while achieving a maximum average diffraction efficiency of $35\%$. We systematically investigate how the diffraction efficiency scales with key grating parameters including transverse size, length, and electron density. Millimeter-scale gratings are shown to sustain diffracted signal intensities up to $2.4\times10^{14}\,\mathrm{W/cm^2}$, demonstrating potential for scalable beam control in multi-petawatt laser systems.

\section{Experimental Setup}
\begin{figure}[t]
    \centering
      \includegraphics[width = \linewidth]{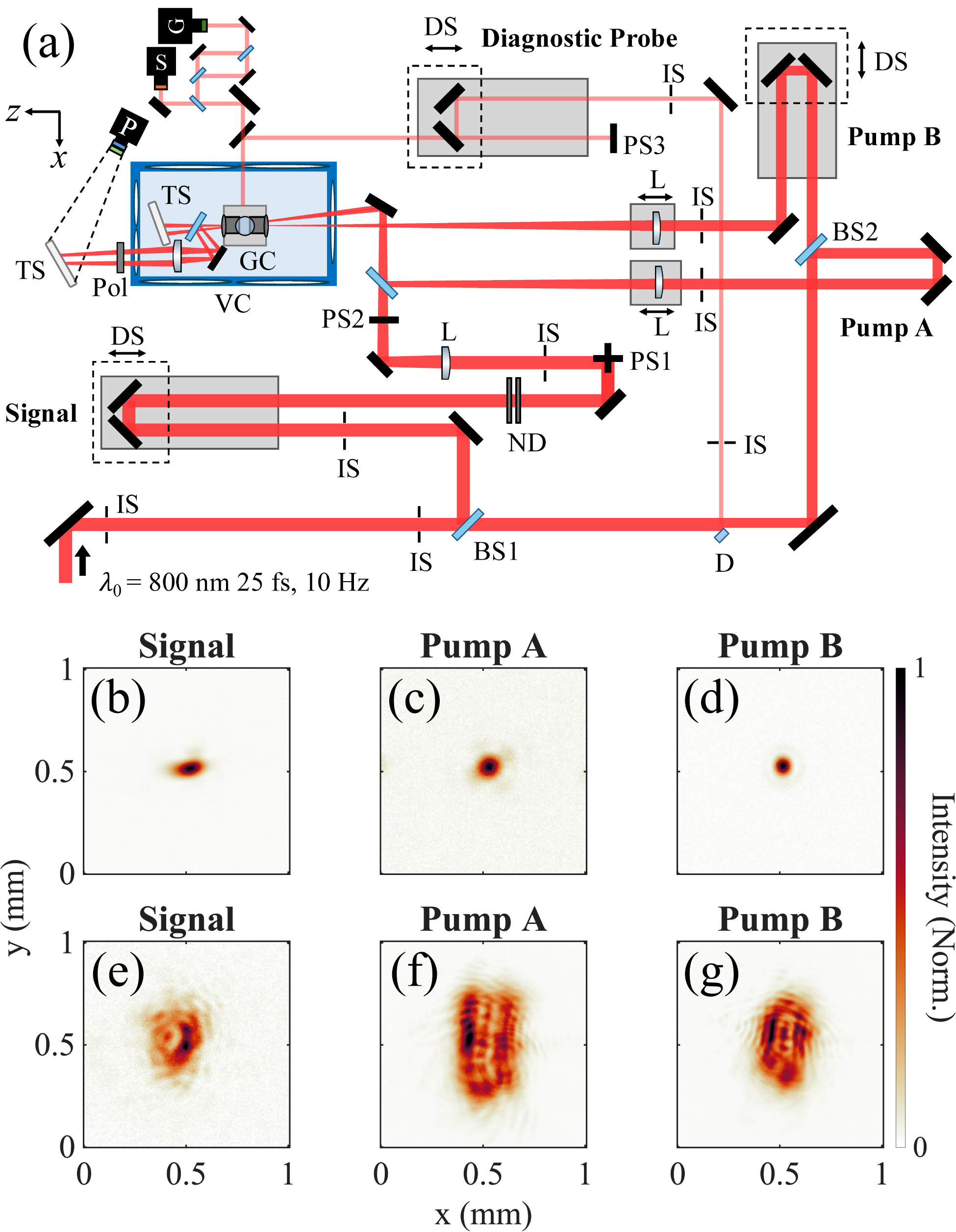}
  \caption{(a) Schematic of the plasma grating experimental setup. Key components: iris (IS), beamsplitter (BS), D-shaped pick-off mirror for a plasma diagnostic probe (D), automated beam shutter (AS), motorized delay stage (DS), neutral density filter (ND), periscope (PS1, PS2, PS3), lens (L), gas cell (GC), vacuum chamber (VC), linear polarizer (Pol), and Teflon scattering screen (TS). Camera P images the pump and signal beam scattering from the Teflon screen to measure the diffraction efficiency, camera S is used for plasma grating shadowgraphy, and camera G for interferometric measurements of the electron density. (b--g) Beam profiles at the gas cell: small-aperture grating -- (b) signal, (c) pump A, (d) pump B; large-aperture grating -- (e) signal, (f) pump A, and (g) pump B.}
    \label{fig:schematic}
\end{figure}

A plasma–neutral grating is generated by spatially and temporally overlapping two femtosecond pump beams in a neutral gas. A schematic of the experimental setup at Princeton is shown in Fig.~\ref{fig:schematic}(a). The pump and signal beams are obtained from a 20-TW Ti:Sapphire laser (Princeton Pulsar 20 TW, Amplitude; central wavelength $\lambda_0 =800 \text{ nm}$, pulse duration $\tau=25 \text{ fs}$, 10-Hz repetition rate). Two s-polarized pump beams (pump A and pump B) propagate in the \textit{xz} plane and intersect at a full crossing angle of $\theta_p=3^\circ$ at the center of a CO$_2$ filled gas cell inside a vacuum chamber, forming a grating with period $\Lambda=15.3~\upmu$m.

Laser input and output apertures are formed by ablating 50-$\upmu$m thick steel foils mounted on the gas cell walls, while maintaining a residual CO$_2$ pressure of $1$--$10$ mbar. After the grating is formed, a high-power signal beam is focused onto the grating at the Bragg angle of incidence:
 \begin{align}
\theta_B = \arcsin{\left(\frac{\lambda_0}{2\Lambda}\right)}
\end{align}
where $\Lambda = {\lambda_1}/\left({2\sin{(\theta_p/2})}\right)$ is the plasma grating period, $\lambda_0$ is the central wavelength of the signal beam, and $\lambda_1$ is the central wavelength of the pumps. For $\lambda_0=\lambda_1 =800$ nm, the Bragg angle is $\theta_B=1.5^\circ$, equal to half the pump crossing angle. The signal beam is p-polarized to avoid interference with the pump beams. Signal and pump beam profiles at the gas cell are shown in Figs. \ref{fig:schematic}(b-d) for the small-aperture grating and Figs. \ref{fig:schematic}(e-g) for the large-aperture grating. After propagation through the gas cell, the undiffracted signal, diffracted signal, and residual pump beams are incident on a Teflon scattering screen (TS), which is imaged with a high dynamic range camera P (Fig.~\ref{fig:schematic}(a)). Beam profiles recorded with  camera P are used to determine the absolute diffraction efficiency $\eta$, defined as the ratio of the diffracted signal counts to the total transmitted signal counts through the gas cell. Absorption of the signal beam in the grating is negligible at studied laser intensities and does not significantly affect the transmission. The diffraction efficiency of the signal beam is measured as a function of the grating dimensions, pump beam energy, electron density $N_e$, and signal beam energy.  

\begin{figure*}
    \centering
      \includegraphics[width = \linewidth]{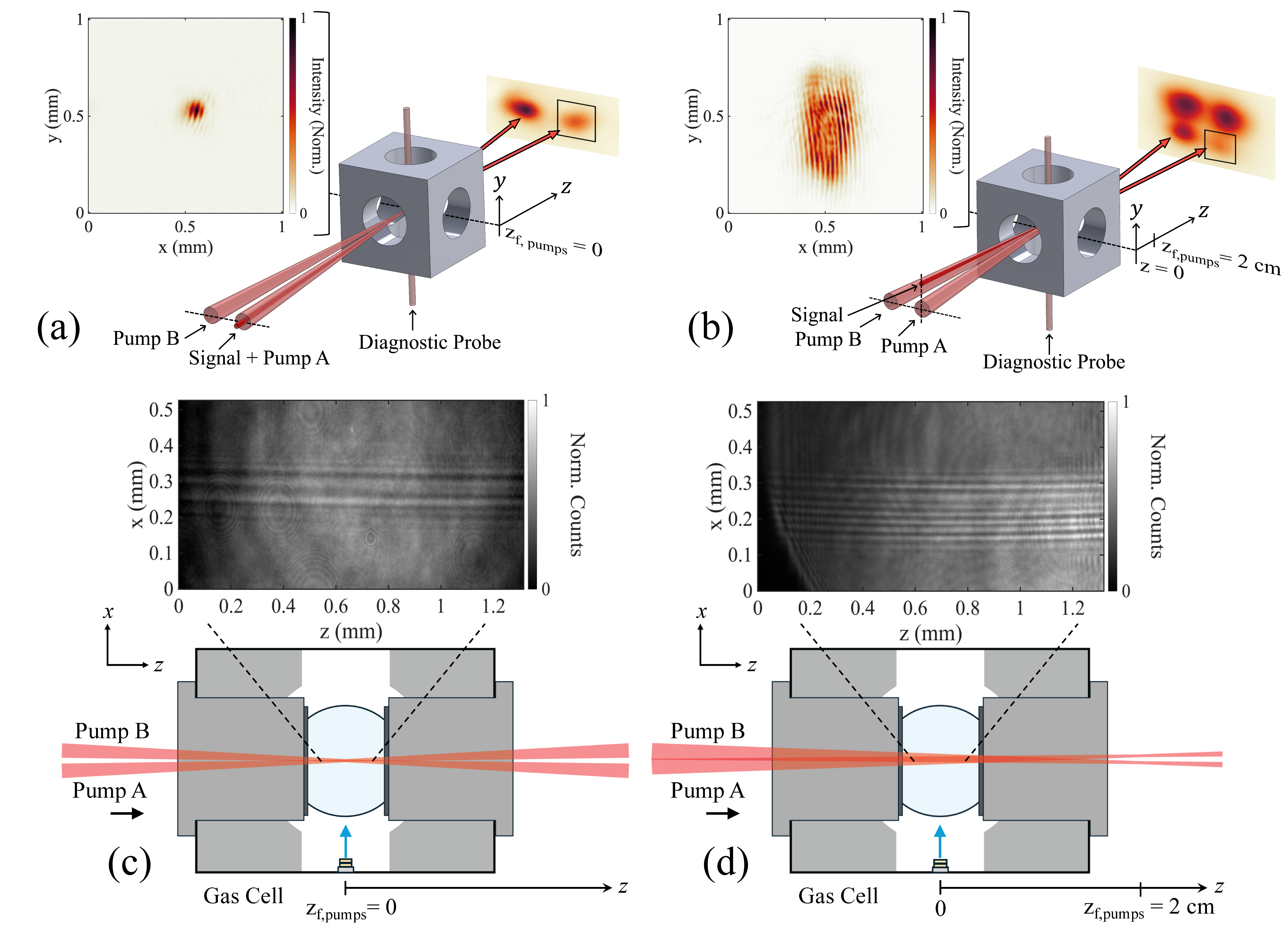}
  \caption{(a) The beam geometry inside the gas cell is shown for the small-aperture grating with pump beams focused at the center of the gas cell at $z_{\text{f,pumps}}=0$ and the signal beam co-linear with pump A. The transverse interference pattern is shown to the left of the GC. The image of the Teflon scattering screen after the gas cell is shown after subtracting out the residual pump beams with the diffracted beam outlined by the black square. (b) The beam geometry for the large-aperture grating with pump beams focused at $z_{\text{f,pumps}} = 2$ cm after the center of the gas cell with the signal beam directly above pump A angled downwards onto the grating. The diffracted signal beam is outlined by the black square. The red arrows indicate the beams' propagation after the gas cell (optics after the gas cell are not depicted). The interference pattern of the pump beams is shown in both (a) and (b) at $z=0$. The top-down view of the pump beams crossing inside the gas cell is shown for the (c) small- and (d) large-aperture grating. Blue arrow denotes direction of gas flow from nozzle. Insets above (c) and (d) show corresponding shadowgraphs.}
  \label{fig:beamgeometry}
\end{figure*}

 \subsection{Beam Geometry for Small- and Large-Aperture Gratings}

\begin{figure}[t]
    \centering
      \includegraphics[width = \linewidth]{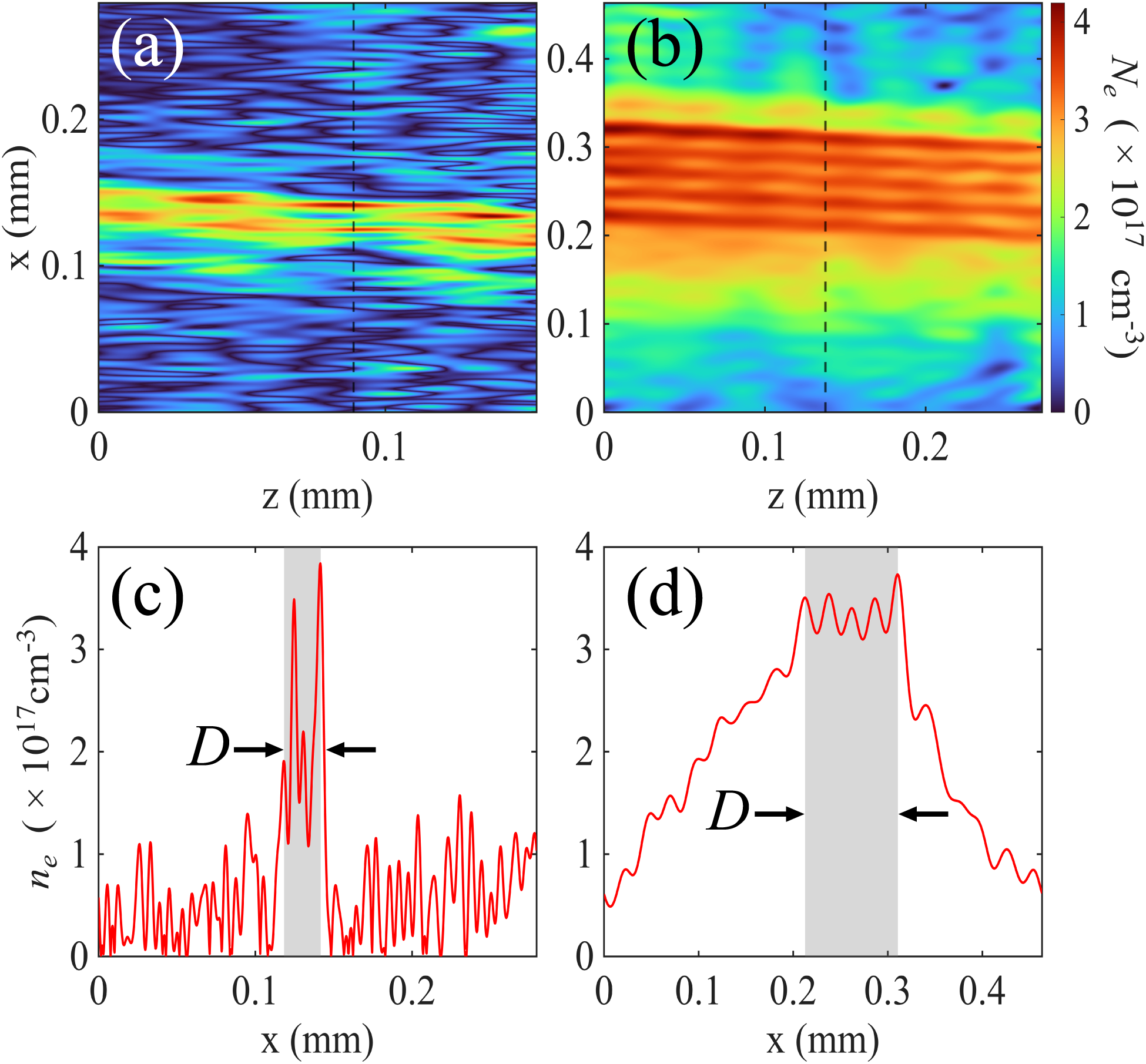}
\caption{Spatial distributions of electron density in the (a)small- and (b) large-aperture plasma grating retrieved by two-dimensional Fourier transform analysis of the plasma-induced phase shift in the interferogram obtained with the folded Mach–Zehnder interferometer. Lineouts of the electron density at $z = 0.09$ mm (small-aperture) and $z = 0.14$ mm (large-aperture) are shown in (c) and (d), respectively. The gray shaded regions indicate the grating apertures, $D = 23$ $\upmu$m and $D = 98$ $\upmu$m. }

\label{fig:shadowgraph}
\end{figure}

Experimental configurations used for grating scalability studies are shown in Fig.~\ref{fig:beamgeometry}. Two plasma gratings of different transverse apertures $D$ were generated and characterized: (1) a small-aperture grating (Fig.~\ref{fig:beamgeometry}(a, c)) with $D=23~\upmu$m generated at the pump focal plane ($z_{\text{f,pumps}}=0$), and (2) a large-aperture grating (Fig.~\ref{fig:beamgeometry}(b, d)) with $D=98~\upmu$m ($4.3\times$ larger) generated $2$ cm before the pump focus ($z_{\text{f,pumps}}=2$ cm), beyond the pump Rayleigh length.  The aperture $D$ was defined as the distance between sharp edges of the transverse electron density profiles, shown by the gray shaded regions in Figs.~\ref{fig:shadowgraph}(b, d). The parameter $z_{\text{f,pumps}}$ in Fig.~\ref{fig:beamgeometry} denotes the position of the pump beam focus relative to the center of the gas cell ($z=0$).

For the small-aperture configuration (Fig.~\ref{fig:beamgeometry}(a)), pump A and pump B propagate in the $x-z$ plane and intersect at their waists at $z_{f,\text{pumps}}=0$. The signal beam (15-mm diameter before focusing) is collinear with pump A and diffracts in the $xz$ plane. After the gas cell, the beams are incident onto a Teflon scattering screen (TS) imaged by camera P (Fig.~\ref{fig:schematic}(a)). For the small-aperture grating setup with in-plane diffraction, a linear polarizer labeled Pol in Fig. \ref{fig:schematic}(a) was placed before the Teflon scattering screen to filter out the residual pump beams since the signal beam was not spatially separated in this configuration. The polarizer was omitted in the large-aperture configuration because the diffracted signal beam was spatially separated as a result of the signal beam being directed out of the horizontal $xz$ plane. The undiffracted and diffracted signal beam profiles on TS are shown in Fig.~\ref{fig:beamgeometry}(a), with the diffracted beam highlighted by the black box after subtraction of the residual pump beam background.

For the large-aperture configuration (Fig.~\ref{fig:beamgeometry}(b)), the pump beams intersect $2$ cm upstream of their foci ($z_{f,\text{pumps}}=2$ cm). The signal beam is directed out of the $xz$ plane by the first periscope (PS1 in Fig. \ref{fig:schematic}(a)) and aligned vertically co-planar with Pump A in the $yz$ plane with azimuthal angle $\theta_B$. It is incident with a downward angle of $\sim2^\circ$ introduced by a second periscope (PS2 in Fig. \ref{fig:schematic}(a)) to spatially separate the undiffracted and diffracted signal beams after interaction with the grating. The diffracted signal beam is outlined by a black box on the Teflon scattering screen behind the gas cell in Fig. \ref{fig:schematic}(b). The signal beam diameter is increased to 19 mm before focusing to enable higher signal energies for damage threshold measurements. The pump interference pattern at $z=0$ is recorded on a camera placed directly at the pumps intersection at alignment power and is shown in the left corner of Figs.~\ref{fig:beamgeometry}(a, b).
A weaker diagnostic probe beam propagating along the $y$ axis transversely through the plasma grating is used for shadowgraphy using camera S (Fig. \ref{fig:schematic}(a)), and interferometric electron density measurements via plasma-induced phase shifts using a folded Mach-Zehnder interferometer \cite{takeda1982fourier} with the grating located outside of the interferometer arms. Plasma grating shadowgraphs and top-view schematics of the pump beam crossing inside the gas cell for the two configurations are shown in Figs. \ref{fig:beamgeometry}(c) and \ref{fig:beamgeometry}(d). Fourier transform analysis of the interferogram recorded by camera G (Fig. \ref{fig:schematic}(a)) is used to extract the electron density profiles of the grating, shown in Fig.~\ref{fig:shadowgraph}.

\section{Scalability and Parameter Dependence of Diffraction Efficiency}

An ideal Bragg grating can reach 100$\%$ diffraction efficiency \cite{moharam1978criterion}. 
However, for focused ultrashort-pulse beams, the spectral bandwidth of the signal and its angular divergence reduce the diffraction efficiency. The resulting diffraction efficiency ($\%$) can be described using coupled-mode theory \cite{yeh1994introduction} as
\begin{equation}
    \eta(L)=\frac{100\%\times\kappa^2}{\kappa^2+\left(\frac{\Delta \alpha }{2}\right)^2}\sin^2\left(\kappa L \left[1+\left(\frac{\Delta \alpha}{2\kappa}\right)^2\right]^\frac{1}{2}\right),
\end{equation}
where $\kappa=\pi n_1/\left(\lambda\cos{\theta_B}\right)$ and $\Delta\alpha=2\pi n_0\left(\cos\theta_2-\cos\theta_1\right)/\lambda$. Here $\theta_1$ and $\theta_2$ are the signal incidence and diffraction angles, $\lambda$ is the wavelength of a spectral component within the signal's bandwidth, $n_0$ is the average refractive index ($n_0\approx1$), $L$ is the grating length, $n_1=n_0-n_\text{plasma}$ is the amplitude of the fundamental Fourier component of the refractive index modulation, and $\kappa L$ is the grating coupling strength \cite{yeh1994introduction}. 
Single-shot Mach–Zehnder interferometry measurements yield peak electron densities between $~1\times 10^{17} \text{ cm}^{-3}$ and $~5\times 10^{17} \text{ cm}^{-3}$. 
Single-shot electron density profiles reconstructed from interferometric phase maps are shown for the small- and large-aperture gratings in Figs.~\ref{fig:shadowgraph}(a) and ~\ref{fig:shadowgraph}(b), with corresponding vertical lineouts in Figs.~\ref{fig:shadowgraph}(c) and ~\ref{fig:shadowgraph}(d), demonstrating well-defined periodic grating structures with apertures $D = 23$ $\upmu$m and $D = 98$ $\upmu$m.

\subsection{Dependence of Diffraction Efficiency on Grating Length}
\begin{figure}[t]
    \centering
      \includegraphics[width = \linewidth]{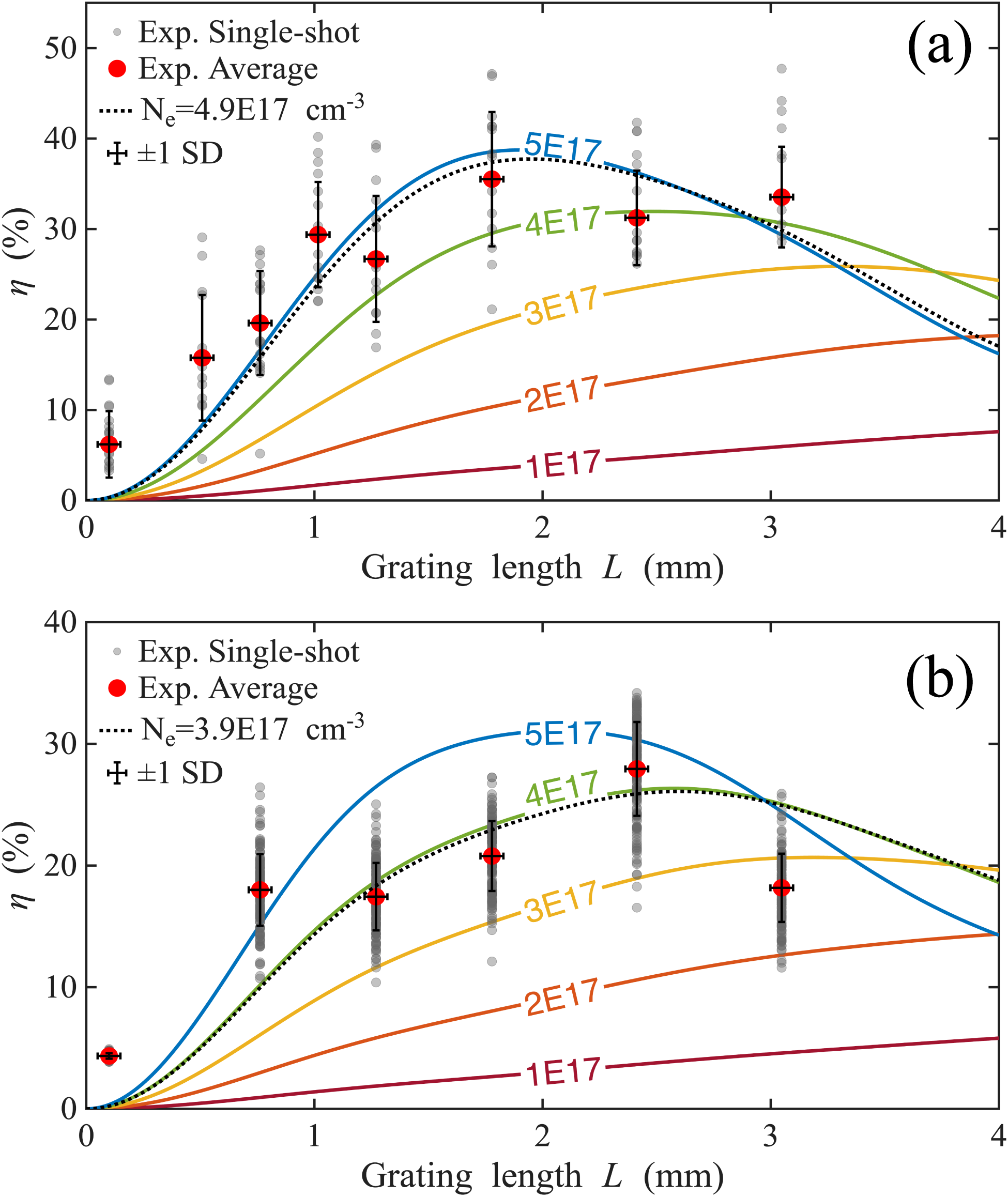}
  \caption{
  Experimental (circles) and theoretical (lines) diffraction efficiency versus grating length for the (a) small- and (b) large-aperture gratings. Red circles show measured average efficiency with black error bars indicating the standard deviation (SD), and gray circles show single-shot efficiencies. The dotted black line is a best-fit obtained by numerical integration of Eq. (2), including Bragg condition mismatch from angular and spectral bandwidths, with the electron density as a free parameter. The fit yields average electron densities of $4.9\times10^{17} \text{cm}^{-3}$ and $3.9\times10^{17} \text{cm}^{-3}$ for the small- and large-aperture gratings, respectively.}
  \label{fig:Diff_eff_vs_length}
\end{figure}
The diffraction efficiency $\eta$ was first characterized as a function of the grating length $L$, controlled by varying the distance between the walls of the gas cell. The measured efficiencies for the small-aperture grating are shown in Fig. \ref{fig:Diff_eff_vs_length}(a) (gray circles denote single shots, red circles denote averages), reaching a maximum average efficiency of $35\%$ and a maximum single-shot efficiency of $48\%$ at $L=$ 1.78 mm. For the large-aperture grating, (Fig. \ref{fig:Diff_eff_vs_length}(b)) the maximum average efficiency is $28\%$ and the maximum single-shot efficiency is $34\%$ at $L =$ 2.41 mm. In both configurations, the efficiency peaks at an optimal grating length and decreases for both shorter and longer gratings.

The solid lines in Fig. \ref{fig:Diff_eff_vs_length}(a, b) show theoretical efficiencies obtained by numerically integrating Eq. (2) for plasma densities between $1\times10^{17}$ and $5\times10^{17}\,\mathrm{cm^{-3}}$, accounting for deviations from the ideal Bragg condition due to the signal beam divergence ($\Delta\theta_1=\pm0.43^\circ$ and $\pm0.54^\circ$ for the small- and large-aperture gratings, respectively) and pulse spectral bandwidth $\Delta\lambda=0.1 \lambda_0=80$ nm. The dotted lines represent fits of Eq. (2) to the experimental data with the electron density $N_e$ as a free parameter, yielding $4.9\times 10^{17} \text{ cm}^{-3}$ and $3.9\times 10^{17} \text{ cm}^{-3}$ for the small- and large-aperture gratings, respectively.
The lower efficiency of the large-aperture grating can be attributed to several factors. First, the larger diameter of the signal beam (19 mm vs 15 mm), used to increase the signal energy for damage-threshold measurements of the large-aperture configuration, increases the angular spread at focus (f-number $f/53$ vs $f/67$ for f=1000 mm lens), leading to greater deviation from the Bragg condition. Second, transverse non-uniformities in the spatial profiles of the pump beams, which, for the large-aperture grating, intersect in their near-field regions before their waists, introduce spatial variations in the interference pattern. Lastly, the average peak plasma density is lower for the large-aperture grating.
\begin{figure}[t]
    \centering
          \includegraphics[width = \linewidth]{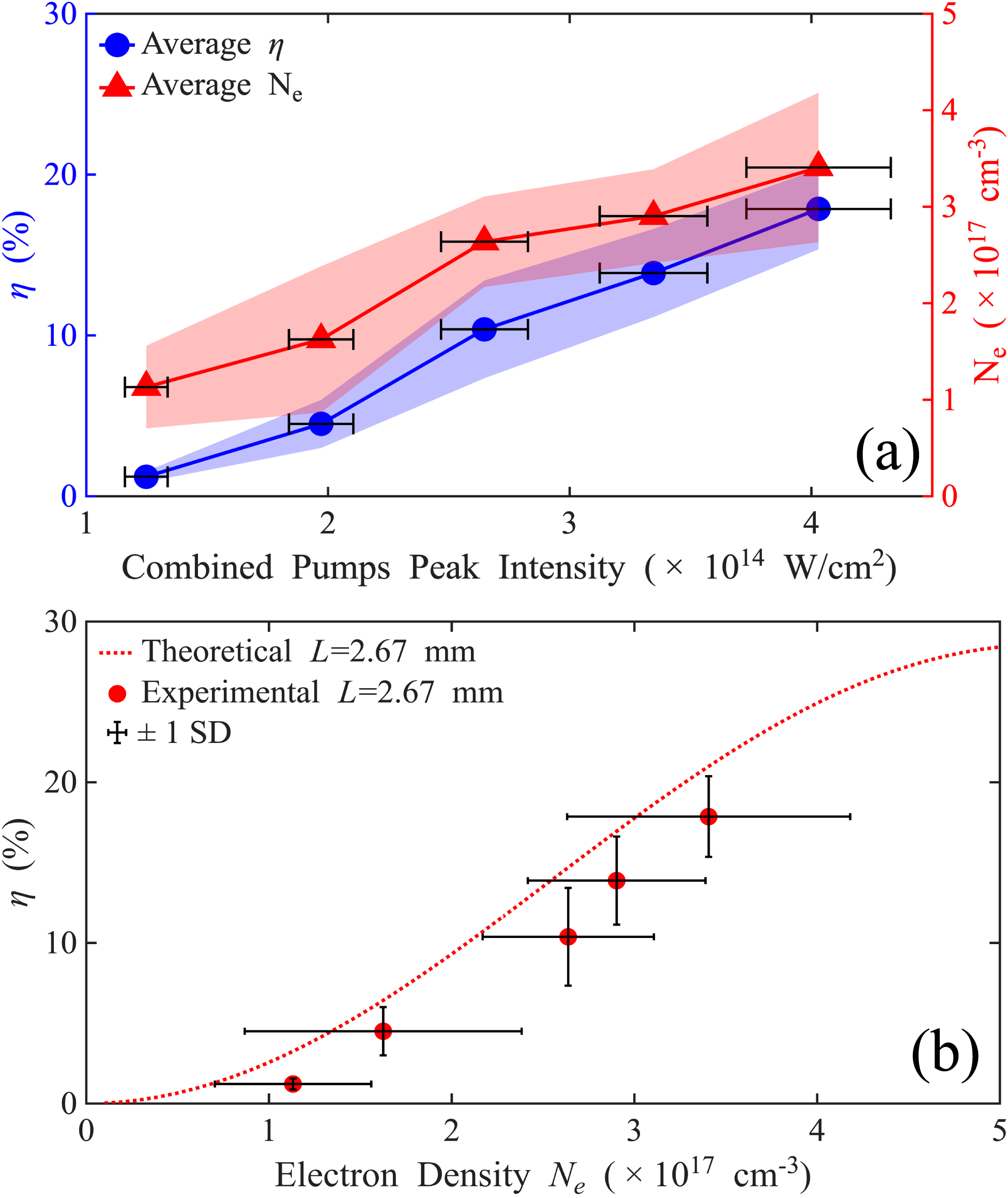}
  \caption{(a) Average diffraction efficiency (blue circles) and electron density (red triangles) as a function of pump intensity for grating length $L=$ 2.67 mm. Blue and red shaded regions indicate the standard deviation (SD) over 400 single shots (efficiency) and 100 single shots (electron density), respectively. Horizontal error bars show the standard deviation of the average intensity of the interference pattern formed by the two pump beams. (b) Average diffraction efficiency as a function of electron density, controlled by varying the pump energy. Red circles show the average experimental $\eta$ for $L=2.67$ mm with error bars corresponding to one standard deviation over 100 single shots. The red dotted line shows the theoretical prediction calculated using $\Delta\lambda=80$ nm and $\Delta\theta_1=\pm0.54^\circ$ in Eq. (2).}
  \label{fig:Diff_eff_pump_energy}
\end{figure}

\subsection{Dependence of Diffraction Efficiency on Pump Intensity and Electron Density}

The dependence of the diffraction efficiency $\eta$ on the pump intensity and corresponding electron density $N_e$, measured with interferometry, was investigated by varying the pump energy for the large-aperture grating. The pump intensity is defined as the average intensity of the interference pattern formed by the two pump beams. The pump intensity is estimated as the intensity of the interference pattern of pump A and pump B, where the intensity of each of the pump beams is given by $I_{A,B}=\sqrt{4\ln2/\pi}\,W_{A,B}/(A_{A,B}\cdot \tau_p)$, where $W_{A,B}$ is the average energy per pulse of pump A and pump B, and $A_{A,B}$ are the cross-sectional areas of the respective pump beams. The pump duration $\tau_p$ is taken as the average of the FWHM intensity durations of pump A and pump B ($29\pm4$ fs), measured with a home-built FROG apparatus. Pump pulse duration and dispersion were adjusted using a dispersive filter (acousto-optic modulator Dazzler, Fastlite) to optimize the diffraction efficiency. The spatial profiles and cross-sectional areas of the pump beams and their interference pattern at the position of the plasma grating were estimated from low-energy beam profiles recorded directly on a CCD, as shown in Figs. \ref{fig:schematic}(c, d, f, g), and Figs.~\ref{fig:beamgeometry}(a, b). Figure~\ref{fig:Diff_eff_pump_energy}(b) shows the measured diffraction efficiency as a function of the measured electron density along with the calculated efficiency (dotted line) using Eq. (2) for a grating length $L=2.67$ mm, taking into account the Bragg condition mismatch due to finite spectral bandwidth, $\Delta\lambda = 80$ nm, and angular divergence ($\Delta\theta_1=\pm0.54^\circ$ about $\theta_B=1.5^\circ$), corresponding to a signal beam of 19 mm diameter incident on an $f$ = 1000 mm lens. These results indicate that higher diffraction efficiencies can be achieved at higher electron densities, corresponding to larger refractive index modulation amplitudes.

\subsection{Intensity Tolerance of Plasma Gratings}
\begin{figure}[t]
    \centering
    \includegraphics[width = \linewidth]{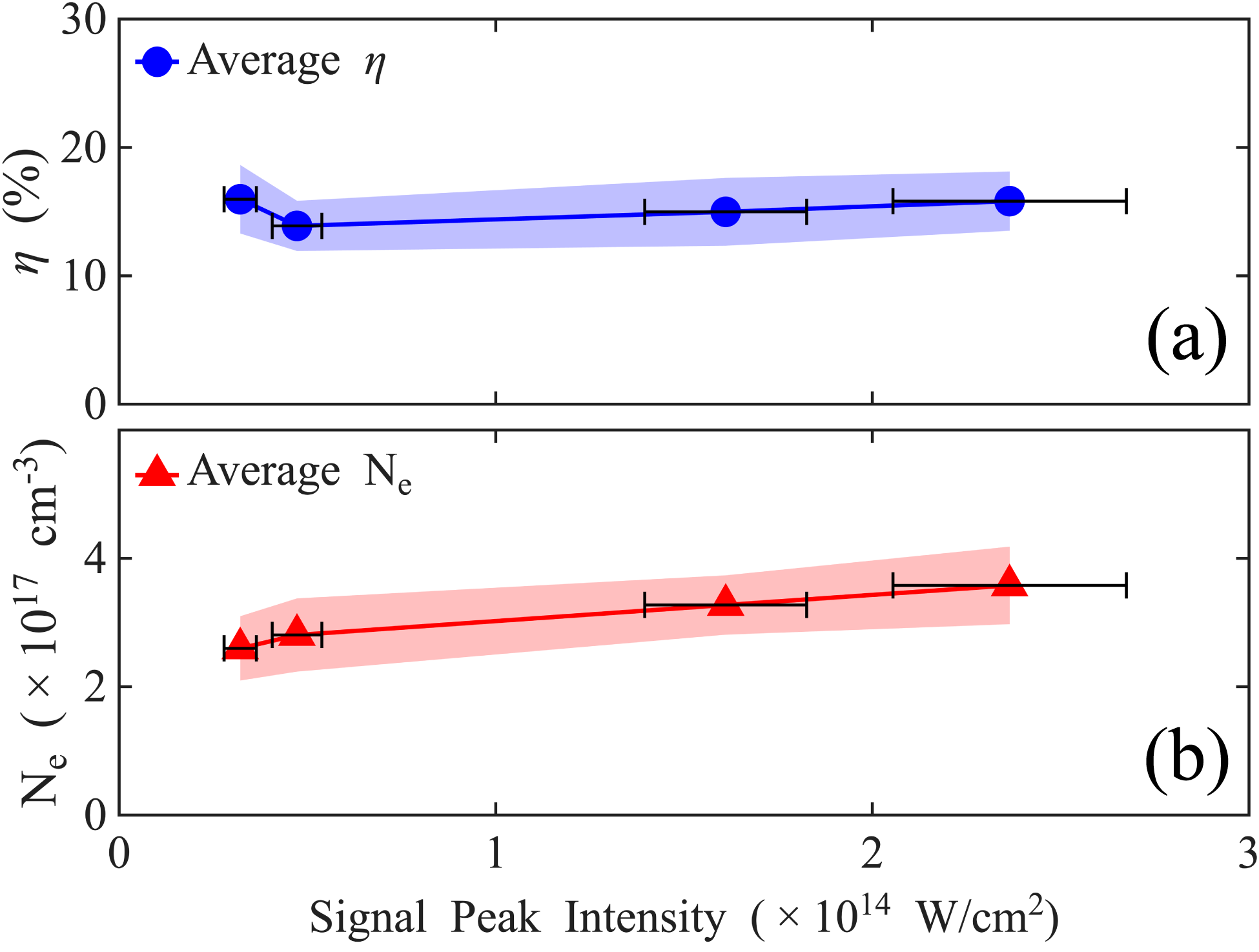}
    \caption{(a) Diffraction efficiency and (b) electron density dependence on signal beam intensity for grating length $L=$ 2.67 mm averaged over 100 single shots. The blue circles show average diffraction efficiency and red triangles show average electron density. Shaded area indicates the standard deviation over 100 single shots and horizontal error bars indicate the standard deviation of signal intensity.}
    \label{fig:ProbeEnergyScan}
\end{figure}

To study the signal-intensity tolerance of the plasma-neutral grating, the signal beam energy was varied from 0.92 mJ to 8.5 mJ using neutral density filters (ND in Fig.~\ref{fig:schematic}). The signal intensity is estimated as $I=\sqrt{4\ln2/\pi}\,W_s/(A_s \tau_s)$, where $W_s$ is the signal beam energy and $A_s$ is the cross-sectional area of the signal beam. The signal duration $\tau_s$ is taken as $41\pm8$ fs, estimated from the FROG measurements taking into account the dispersion in transmissive optics. The spatial profile (Fig.~\ref{fig:schematic}(b, e)) and cross-sectional area of the signal beam were recorded directly on a CCD placed at the position of the plasma grating with an attenuated signal beam. The peak signal intensities at the plasma grating varied from $\sim 3.2\times10^{13}\text{ W/cm}^2$ to $2.4\times10^{14}\text{ W/cm}^2$ (Fig.~\ref{fig:ProbeEnergyScan}). Across this range of signal intensities, the diffraction efficiency decreased by less than 1\% (Fig.~\ref{fig:ProbeEnergyScan}(a)), indicating no measurable degradation of the grating on the interaction timescale.
The average peak electron density increased by 38\%, from $2.6\times10^{17}$ to $3.6\times10^{17}\,\text{ cm}^{-3}$  (Fig.~\ref{fig:ProbeEnergyScan}(b)). These results indicate weak signal-induced ionization without measurable degradation of grating performance. Long term stability was further evaluated by continuous operation for one hour at 10 Hz pulse repetition rate, during which the diffraction efficiency fluctuated by 2\% (standard deviation) with no systematic decrease.

\section{Plasma Grating Lifetimes}

The lifetime of the ionization-induced refractive index modulation sets the operational time window of plasma–neutral gratings for high-intensity and multi-petawatt laser applications. The evolution and lifetime of the transient plasma-neutral grating were characterized by scanning a temporal delay of the signal pulse relative to the pump pulses, denoted by $\Delta t_{signal}$, and monitoring the diffraction efficiency of the signal. In molecular gases, such as CO$_2$ used in this work, both collisional electron recombination and diffusion \cite{zhang2021ionization, jarnac2014study} contribute to the degradation of the grating structure, while in atomic gases the dominant relaxation mechanism is the ambipolar diffusion \cite{durand2012dynamics,jarnac2014study}. The plasma structure decay also depends on the grating period. Smaller periods accelerate washout of the refractive index modulation, and the grating lifetime becomes shorter than the characteristic electron recombination time, and is therefore primarily diffusion-limited \cite{jarnac2014study}. 

\begin{figure}[t]
    \centering
    \begin{tabular}{cc}
      \includegraphics[width = \linewidth]{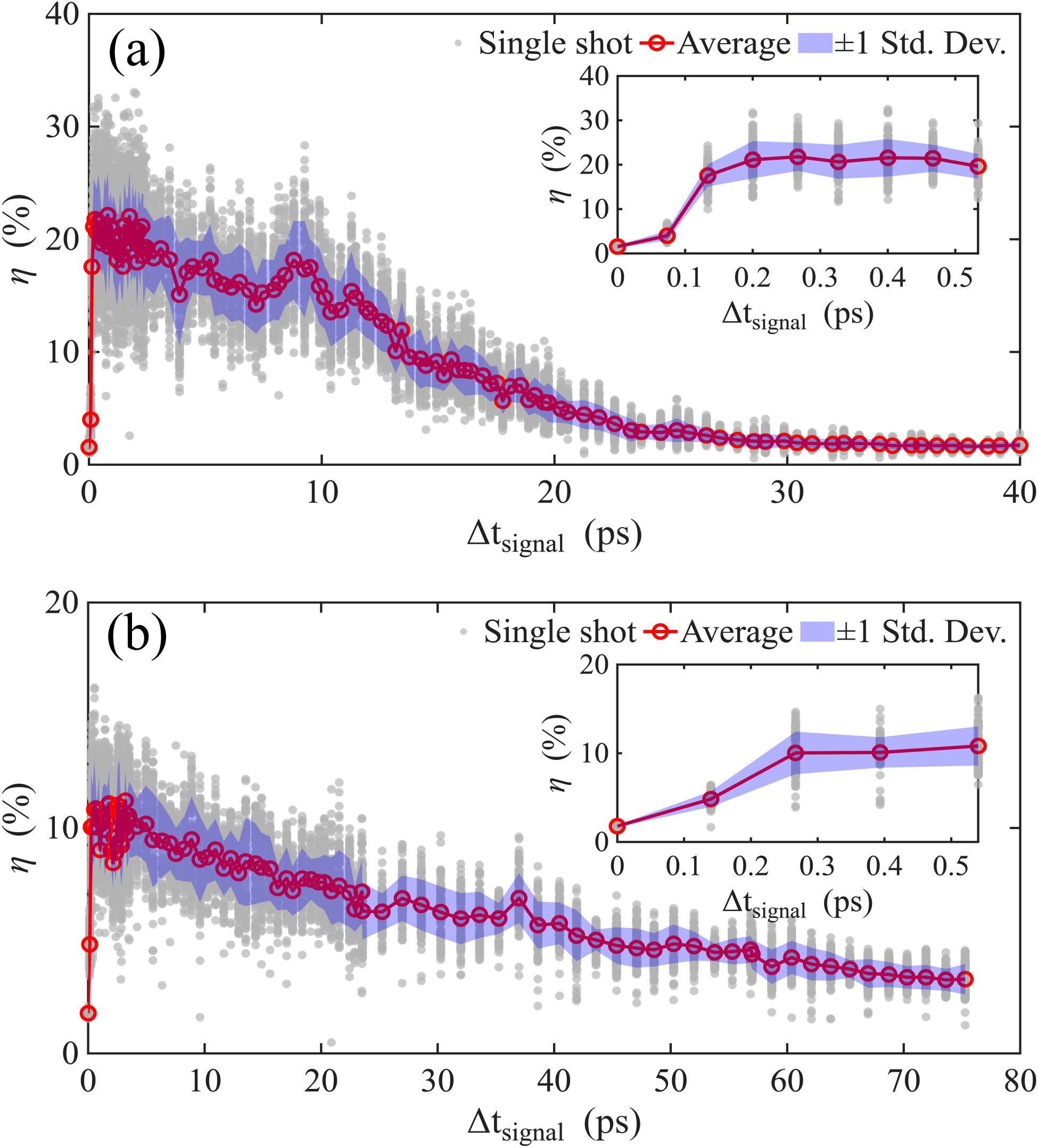}
    \end{tabular}
  \caption{Diffraction efficiency as a function of the pump-signal delay  $\Delta t_{\text{signal}}$ for the (a) small- and (b) large-aperture grating. Gray circles show each of the 100 single shots taken at each signal delay and red circles show the average of each burst of 100 individual shots. The shaded blue region indicates one standard deviation. Maximum single shot diffraction efficiency for the small- and large-aperture gratings are $33\%$ and $16\%$ respectively. Inset plots show grating switch-on behavior immediately after $\Delta t_{\text{signal}}=0$.}
  \label{fig:ProbeDelay}
\end{figure}

In this work, the plasma grating structure remained stable without significant degradation for tens of picoseconds in both small- and large-aperture configurations, as is demonstrated by the pump-signal delay scans of the diffraction efficiency shown in Fig.~\ref{fig:ProbeDelay}(a, b), where gray circles denote 100 single-shot measurements at each delay and red circles show the averages. Zero delay corresponds to the maximum temporal overlap between the signal and pump pulses. Positive signal delay indicates the signal pulse arriving later in time relative to the pump pulses. Exponential fits to the data yield $1/e$ lifetimes of $\tau_{1/e}=17$ ps and $\tau_{1/e}=68$ ps for the small- and large-aperture gratings, respectively. The longer lifetime and slower switch-on time of the large-aperture grating may arise from spatial-temporal coupling where the interference, produced by the pump beams in their near-fields, has a larger effective interaction region. In contrast, the interference pattern for the small-aperture grating is more tightly confined in space and time, resulting in faster switch-on and decay outside the temporal overlap. The observed picosecond-scale lifetime of the refractive index modulation in the plasma grating structures support efficient diffraction of ultrashort pulses.

\section{Conclusion}
To summarize, we used the basic physics of strong-field light-matter interaction to create reconfigurable plasma-based photonic structures in a neutral gas for controllable diffraction of high-intensity ultrashort laser pulses. and characterized their performance and scalability. The diffraction efficiency of the plasma gratings was measured as a function of grating dimensions, pump laser intensity, electron density, and diffracted signal intensity. Two configurations were under study: a small-aperture grating, formed by the focused pump beams crossing at their waists, achieving up to 35\% average diffraction efficiency, and a larger aperture grating formed by the pump beams crossing before their waists, capable of diffracting higher-energy signal pulses with average efficiencies up to 28\%. For both configurations, the diffraction efficiency was optimal at a specific grating length in agreement with the coupled-mode theory for Bragg diffraction. The grating lifetime, measured by scanning the pump-signal delay, is on the order of 10 ps, which is long enough to control ultrashort pulses. We observed stable continuous operation of plasma gratings at signal laser intensities of up to $2.4\times10^{14}$ W/cm$^2$ (fluence $\approx 10.3$ J/cm$^2$), substantially exceeding the damage threshold of solid-state optics. The efficiency and stability of large-aperture gratings demonstrate the scalability of plasma gratings for applications  requiring larger plasma optic sizes such as a diffractive plasma lens or plasma grating for higher-energy lasers. Further improvements in diffraction efficiency can be achieved through laser beam pointing stabilization, correction of pump wavefront aberrations using adaptive optics, reduction of Bragg condition mismatch, and optimization of the grating coupling strength $\kappa L$. These results demonstrate the feasibility of transient plasma photonic structures as scalable optical components for high-power laser systems.

\begin{acknowledgments}
This work was partially supported by the U.S. Department of Energy under grants No. DE-SC0025497, DE-SC002557, the U.S. National Science Foundation under grants No. PHY-2206711, PHY-2308641, PHY-2512131, and PHY-2541940, the National Science Foundation Graduate Research Fellowship under grant No. DGE-2146755 (V. M. P-R.), the National Nuclear Security Administration under grant No. DE-NA0004130, and the Gordon and Betty Moore Foundation Grant DOI 10.37807/gbmf12255. Lawrence Livermore National Laboratory is operated by Lawrence Livermore National Security, LLC, for the U.S. Department of Energy, National Nuclear Security Administration under Contract No. DE-AC52-07NA27344.

\end{acknowledgments}

\bibliography{References}

@inproceedings{Mikhailova2025,
  title = {Plasma optics for ultrafast high-intensity light sources},
  url = {http://dx.doi.org/10.1117/12.3065349},
  DOI = {10.1117/12.3065349},
  booktitle = {Coherent and Incoherent Radiation Sources based on Laser-driven Relativistic Plasma Waves},
  publisher = {SPIE},
  author = {Mikhailova,  Julia},
  editor = {Jaroszynski,  Dino A. and Hur,  MinSup},
  year = {2025},
  month = jun,
  pages = {13}
}

@phdthesis{fasano2023harmonic,
  title={Harmonic Generation in Reflection From Plasma Mirrors},
  author={Fasano, Nicholas Maurice},
  year={2023},
  school={Princeton University},
  note = {Available at http://arks.princeton.edu/ark:/88435/dsp01rn3014630}
}

@article{Mikhailova2012,
  title={Isolated attosecond pulses from laser-driven synchrotron radiation},
  author={Mikhailova, Julia M and Fedorov, MV and Karpowicz, Nicholas and Gibbon, Paul and Platonenko, VT and Zheltikov, AM and Krausz, Ferenc},
  journal={Physical Review Letters},
  volume={109},
  number={24},
  pages={245005},
  year={2012},
  publisher={APS}
}

@article{edwards2016multipass,
  title={Multipass relativistic high-order-harmonic generation for intense attosecond pulses},
  author={Edwards, Matthew R and Mikhailova, Julia M},
  journal={Physical Review A},
  volume={93},
  number={2},
  pages={023836},
  year={2016},
  publisher={APS}
}

@article{Jahn2019,
  title = {Towards intense isolated attosecond pulses from relativistic surface high harmonics},
  volume = {6},
  ISSN = {2334-2536},
  url = {http://dx.doi.org/10.1364/OPTICA.6.000280},
  DOI = {10.1364/optica.6.000280},
  number = {3},
  journal = {Optica},
  publisher = {Optica Publishing Group},
  author = {Jahn,  Olga and Leshchenko,  Vyacheslav E. and Tzallas,  Paraskevas and Kessel,  Alexander and Kr\"{u}ger,  Mathias and M\"{u}nzer,  Andreas and Trushin,  Sergei A. and Tsakiris,  George D. and Kahaly,  Subhendu and Kormin,  Dmitrii and Veisz,  Laszlo and Pervak,  Vladimir and Krausz,  Ferenc and Major,  Zsuzsanna and Karsch,  Stefan},
  year = {2019},
  month = mar,
  pages = {280}
}

@article{Heissler2012,
  title={Few-Cycle Driven Relativistically Oscillating Plasma Mirrors: A Source of Intense Isolated Attosecond Pulses},
  author={Heissler, Patrick and H{\"o}rlein, Rainer and Mikhailova, Julia M and Waldecker, Lutz and Tzallas, P and Buck, Alexander and Schmid, Karl and Sears, CMS and Krausz, Ferenc and Veisz, L{\'a}szl{\'o} and others},
  journal={Physical Review Letters},
  volume={108},
  number={23},
  pages={235003},
  year={2012},
  publisher={APS}
}

@article{Haessler2022,
  title={High-harmonic generation and correlated electron emission from relativistic plasma mirrors at 1 \text{kHz} repetition rate},
  author={Haessler, Stefan and Ouill{\'e}, Marie and Kaur, Jaismeen and Bocoum, Ma{\"\i}mouna and B{\"o}hle, Frederik and Levy, Dan and Daniault, Louis and Vernier, Aline and Faure, J{\'e}r{\^o}me and Lopez-Martens, Rodrigo},
  journal={Ultrafast Science},
  year={2022},
  publisher={AAAS}
}

@article{Chopineau2022,
  title={Sub-laser-cycle control of relativistic plasma mirrors},
  author={Chopineau, Ludovic and Blaclard, Guillaume and Denoeud, Adrien and Vincenti, Henri and Qu{\'e}r{\'e}, Fabien and Haessler, Stefan},
  journal={Physical Review Research},
  volume={4},
  number={1},
  pages={L012030},
  year={2022},
  publisher={APS}
}

@article{Ouill2024,
  title = {Lightwave-controlled relativistic plasma mirrors},
  volume = {49},
  ISSN = {1539-4794},
  url = {http://dx.doi.org/10.1364/OL.534255},
  DOI = {10.1364/ol.534255},
  number = {17},
  journal = {Optics Letters},
  publisher = {Optica Publishing Group},
  author = {Ouillé,  Marie and Kaur,  Jaismeen and Cheng,  Zhao and Haessler,  Stefan and Lopez-Martens,  Rodrigo},
  year = {2024},
  month = aug,
  pages = {4847}
}

@article{Fasano2023,
  title={Electron bunch dynamics and emission in particle-in-cell simulations of relativistic laser--solid interactions: On density artifacts, collisions, and numerical dispersion},
  author={Fasano, Nicholas M and Edwards, Matthew R and Mikhailova, Julia M},
  journal={Physics of Plasmas},
  volume={30},
  number={6},
  year={2023},
  publisher={AIP Publishing}
}

@article{Edwards2020,
  title={Electron-nanobunch-width-dominated spectral power law for relativistic harmonic generation from ultrathin foils},
  author={Edwards, Matthew R and Fasano, Nicholas M and Mikhailova, Julia M},
  journal={Physical Review Letters},
  volume={124},
  number={18},
  pages={185004},
  year={2020},
  publisher={APS}
}

@article{Leshchenko2019,
  title={On-target temporal characterization of optical pulses at relativistic intensity},
  author={Leshchenko, Vyacheslav E and Kessel, Alexander and Jahn, Olga and Kr{\"u}ger, Mathias and M{\"u}nzer, Andreas and Trushin, Sergei A and Veisz, Laszlo and Major, Zsuzsanna and Karsch, Stefan},
  journal={Light: Science \& Applications},
  volume={8},
  number={1},
  pages={96},
  year={2019},
  publisher={Nature Publishing Group UK London}
}

@article{edwards2014enhanced,
  title={Enhanced attosecond bursts of relativistic high-order harmonics driven by two-color fields},
  author={Edwards, Matthew R and Platonenko, Victor T and Mikhailova, Julia M},
  journal={Optics Letters},
  volume={39},
  number={24},
  pages={6823--6826},
  year={2014},
  publisher={Optical Society of America}
}

@article{Morozov2018,
  title={Ionization assisted self-guiding of femtosecond laser pulses},
  author={Morozov, A and Goltsov, A and Chen, Q and Scully, M and Suckewer, S},
  journal={Physics of Plasmas},
  volume={25},
  number={5},
  year={2018},
  publisher={AIP Publishing}
}

@article{Feder2020,
  title={Self-waveguiding of relativistic laser pulses in neutral gas channels},
  author={Feder, L and Miao, B and Shrock, JE and Goffin, A and Milchberg, HM},
  journal={Physical Review Research},
  volume={2},
  number={4},
  pages={043173},
  year={2020},
  publisher={APS}
}

@article{Picksley2024,
  title={Matched guiding and controlled injection in dark-current-free, 10-\text{GeV}-class, channel-guided laser-plasma accelerators},
  author={Picksley, A and Stackhouse, J and Benedetti, C and Nakamura, K and Tsai, HE and Li, R and Miao, B and Shrock, JE and Rockafellow, E and Milchberg, HM and others},
  journal={Physical Review Letters},
  volume={133},
  number={25},
  pages={255001},
  year={2024},
  publisher={APS}
}

@article{Rankin1991,
  title = {Refraction effects associated with multiphoton ionization and ultrashort-pulse laser propagation in plasma waveguides},
  volume = {16},
  ISSN = {1539-4794},
  url = {http://dx.doi.org/10.1364/ol.16.000835},
  DOI = {10.1364/ol.16.000835},
  number = {11},
  journal = {Optics Letters},
  publisher = {Optica Publishing Group},
  author = {Rankin,  R. and Burnett,  N. H. and Corkum,  P. B. and Capjack,  C. E.},
  year = {1991},
  month = jun,
  pages = {835}
}

@article{DiPiazza2012,
  title = {Extremely high-intensity laser interactions with fundamental quantum systems},
  volume = {84},
  ISSN = {1539-0756},
  url = {http://dx.doi.org/10.1103/RevModPhys.84.1177},
  DOI = {10.1103/revmodphys.84.1177},
  number = {3},
  journal = {Reviews of Modern Physics},
  publisher = {American Physical Society (APS)},
  author = {Di Piazza,  A. and M\"{u}ller,  C. and Hatsagortsyan,  K. Z. and Keitel,  C. H.},
  year = {2012},
  month = aug,
  pages = {1177–1228}
}

@article{Los2026,
  title={Observation of quantum effects on radiation reaction in strong fields},
  author={Los, Eva E and Gerstmayr, Elias and Arran, Christopher and Streeter, Matthew JV and Colgan, Cary and Cobo, Claudia C and Kettle, Brendan and Blackburn, Thomas G and Bourgeois, Nicolas and Calvin, Luke and others},
  journal={Nature Communications},
  year={2026},
  publisher={Nature Publishing Group UK London}
}

@article{Yoon2021,
  title = {Realization of laser intensity over $10^{23}$ {W}/cm$^2$},
  volume = {8},
  ISSN = {2334-2536},
  url = {http://dx.doi.org/10.1364/OPTICA.420520},
  DOI = {10.1364/optica.420520},
  number = {5},
  journal = {Optica},
  publisher = {Optica Publishing Group},
  author = {Yoon,  Jin Woo and Kim,  Yeong Gyu and Choi,  Il Woo and Sung,  Jae Hee and Lee,  Hwang Woon and Lee,  Seong Ku and Nam,  Chang Hee},
  year = {2021},
  month = may,
  pages = {630}
}

@article{Kramer2023,
  title = {{New SLAC x-ray laser fires its first photons}},
  volume = {2023},
  ISSN = {1945-0699},
  url = {http://dx.doi.org/10.1063/PT.6.2.20230918a},
  number = {09},
  journal = {Physics Today},
  publisher = {AIP Publishing},
  author = {Kramer,  David},
  year = {2023},
  month = sep 
}

@article{Guo2018,
  title = {Improvement of the focusing ability by double deformable mirrors for 10 {PW}-level {T}i:{S}apphire chirped pulse amplification laser system},
  volume = {26},
  ISSN = {1094-4087},
  url = {http://dx.doi.org/10.1364/OE.26.026776},
  DOI = {10.1364/oe.26.026776},
  number = {20},
  journal = {Optics Express},
  publisher = {Optica Publishing Group},
  author = {Guo,  Zhen and Yu,  Lianghong and Wang,  Jianye and Wang,  Cheng and Liu,  Yanqi and Gan,  Zebiao and Li,  Wenqi and Leng,  Yuxin and Liang,  Xiaoyan and Li,  Ruxin},
  year = {2018},
  month = sep,
  pages = {26776}
}

@article{Pirozhkov2017,
  title = {Approaching the diffraction-limited,  bandwidth-limited Petawatt},
  volume = {25},
  ISSN = {1094-4087},
  url = {http://dx.doi.org/10.1364/OE.25.020486},
  DOI = {10.1364/oe.25.020486},
  number = {17},
  journal = {Optics Express},
  publisher = {Optica Publishing Group},
  author = {Pirozhkov,  Alexander S. and Fukuda,  Yuji and Nishiuchi,  Mamiko and Kiriyama,  Hiromitsu and Sagisaka,  Akito and Ogura,  Koichi and Mori,  Michiaki and Kishimoto,  Maki and Sakaki,  Hironao and Dover,  Nicholas P. and Kondo,  Kotaro and Nakanii,  Nobuhiko and Huang,  Kai and Kanasaki,  Masato and Kondo,  Kiminori and Kando,  Masaki},
  year = {2017},
  month = aug,
  pages = {20486}
}

@article{Bromage2019,
  title = {Technology development for ultraintense all-{OPCPA} systems},
  volume = {7},
  ISSN = {2052-3289},
  url = {http://dx.doi.org/10.1017/hpl.2018.64},
  journal = {High Power Laser Science and Engineering},
  publisher = {Cambridge University Press (CUP)},
  author = {Bromage,  J. and Bahk,  S.-W. and Begishev,  I. A. and Dorrer,  C. and Guardalben,  M. J. and Hoffman,  B. N. and Oliver,  J. B. and Roides,  R. G. and Schiesser,  E. M. and Shoup III,  M. J. and Spilatro,  M. and Webb,  B. and Weiner,  D. and Zuegel,  J. D.},
  year = {2019}
}

@article{Veisz2025,
  title = {{Waveform-controlled field synthesis of sub-two-cycle pulses at the 100 TW peak power level}},
  volume = {19},
  ISSN = {1749-4893},
  url = {http://dx.doi.org/10.1038/s41566-025-01720-2},
  DOI = {10.1038/s41566-025-01720-2},
  number = {9},
  journal = {Nature Photonics},
  publisher = {Springer Science and Business Media LLC},
  author = {Veisz,  Laszlo and Fischer,  Peter and Vardast,  Sajjad and Schnur,  Fritz and Muschet,  Alexander and De Andres,  Aitor and Kaniyeri,  Sreehari and Li,  Hang and Salh,  Roushdey and Ferencz,  Kárpát and Nagy,  Gergely Norbert and Kahaly,  Subhendu},
  year = {2025},
  month = jul,
  pages = {1013–1019}
}

@article{Gales2018,
  title = {{The extreme light infrastructure—nuclear physics (ELI-NP) facility: new horizons in physics with 10 PW ultra-intense lasers and 20 MeV brilliant gamma beams}},
  volume = {81},
  ISSN = {1361-6633},
  url = {http://dx.doi.org/10.1088/1361-6633/aacfe8},
  DOI = {10.1088/1361-6633/aacfe8},
  number = {9},
  journal = {Reports on Progress in Physics},
  publisher = {IOP Publishing},
  author = {Gales,  S and Tanaka,  K A and Balabanski,  D L and Negoita,  F and Stutman,  D and Tesileanu,  O and Ur,  C A and Ursescu,  D and Andrei,  I and Ataman,  S and Cernaianu,  M O and D’Alessi,  L and Dancus,  I and Diaconescu,  B and Djourelov,  N and Filipescu,  D and Ghenuche,  P and Ghita,  D G and Matei,  C and Seto,  K and Zeng,  M and Zamfir,  N V},
  year = {2018},
  month = aug,
  pages = {094301}
}

@article{AbuShawareb2024,
  title = {Achievement of Target Gain Larger than Unity in an Inertial Fusion Experiment},
  volume = {132},
  ISSN = {1079-7114},
  url = {http://dx.doi.org/10.1103/PhysRevLett.132.065102},
  number = {6},
  journal = {Physical Review Letters},
  publisher = {American Physical Society (APS)},
  author = {Abu-Shawareb,  H. and Acree,  R. and Adams,  P. and Adams,  J. and Addis,  B. and Aden,  R. and Adrian,  P. and Afeyan, B. and Aggleton, M and et al},
  year = {2024},
  month = feb 
}

@article{Wang2017ALEPH,
  title = {0.85 {PW} laser operation at 33 {H}z and high-contrast ultrahigh-intensity $\lambda$ =400 nm second-harmonic beamline},
  volume = {42},
  ISSN = {1539-4794},
  url = {http://dx.doi.org/10.1364/OL.42.003828},
  DOI = {10.1364/ol.42.003828},
  number = {19},
  journal = {Optics Letters},
  publisher = {Optica Publishing Group},
  author = {Wang,  Yong and Wang,  Shoujun and Rockwood,  Alex and Luther,  Bradley M. and Hollinger,  Reed and Curtis,  Alden and Calvi,  Chase and Menoni,  Carmen S. and Rocca,  Jorge J.},
  year = {2017},
  month = sep,
  pages = {3828}
}

@article{Zheltikov2024_PhysLettA,
  title = {Wait time to stochastic self-focusing},
  volume = {505},
  ISSN = {0375-9601},
  url = {http://dx.doi.org/10.1016/j.physleta.2024.129432},
  DOI = {10.1016/j.physleta.2024.129432},
  journal = {Physics Letters A},
  publisher = {Elsevier BV},
  author = {Zheltikov,  A.M.},
  year = {2024},
  month = may,
  pages = {129432}
}

@article{Zheltikov2024_PRA,
  title = {Criteria for stochastic self-focusing},
  volume = {110},
  ISSN = {2469-9934},
  url = {http://dx.doi.org/10.1103/PhysRevA.110.023516},
  number = {2},
  journal = {Physical Review A},
  publisher = {American Physical Society (APS)},
  author = {Zheltikov,  A. M.},
  year = {2024},
  month = aug 
}

@article{Edwards2020_Two_color,
  title = {A multi-terawatt two-color beam for high-power field-controlled nonlinear optics},
  volume = {45},
  ISSN = {1539-4794},
  url = {http://dx.doi.org/10.1364/OL.403806},
  DOI = {10.1364/ol.403806},
  number = {23},
  journal = {Optics Letters},
  publisher = {Optica Publishing Group},
  author = {Edwards,  M. R. and Fasano,  N. M. and Bennett,  T. and Griffith,  A. and Turley,  N. and O’Brien,  B. M. and Mikhailova,  J. M.},
  year = {2020},
  month = nov,
  pages = {6542}
}

@article{Mikhailova2011,
  title = {Ultra-high-contrast few-cycle pulses for multipetawatt-class laser technology},
  volume = {36},
  ISSN = {1539-4794},
  url = {http://dx.doi.org/10.1364/OL.36.003145},
  DOI = {10.1364/ol.36.003145},
  number = {16},
  journal = {Optics Letters},
  publisher = {Optica Publishing Group},
  author = {Mikhailova,  Julia M. and Buck,  Alexander and Borot,  Antonin and Schmid,  Karl and Sears,  Christopher and Tsakiris,  George D. and Krausz,  Ferenc and Veisz,  Laszlo},
  year = {2011},
  month = aug,
  pages = {3145}
}

@article{Edwards2017,
  title={X-ray amplification by stimulated Brillouin scattering},
  author={Edwards, Matthew R and Mikhailova, Julia M and Fisch, Nathaniel J},
  journal={Physical Review E},
  volume={96},
  number={2},
  pages={023209},
  year={2017},
  publisher={APS}
}

@article{edwards2022plasma,
  title={Plasma transmission gratings for compression of high-intensity laser pulses},
  author={Edwards, Matthew R and Michel, Pierre},
  journal={Physical Review Applied},
  volume={18},
  number={2},
  pages={024026},
  year={2022},
  publisher={APS}
}

@article{edwards2024greater,
  title={Greater than five-order-of-magnitude postcompression temporal contrast improvement with an ionization plasma grating},
  author={Edwards, Matthew R and Fasano, Nicholas M and Giakas, Andreas M and Wang, Michelle M and Griff-McMahon, Jesse and Morozov, Anatoli and Perez-Ramirez, Victor M and Lemos, Nuno and Michel, Pierre and Mikhailova, Julia M},
  journal={Physical Review Letters},
  volume={133},
  number={15},
  pages={155101},
  year={2024},
  publisher={APS}
}

@article{edwards2022holographic,
  title={Holographic plasma lenses},
  author={Edwards, M. R. and Munirov, V. R and Singh, A and Fasano, N. M. and Kur, E and Lemos, N and Mikhailova, J. M. and Wurtele, J. S. and Michel, P},
  journal={Physical Review Letters},
  volume={128},
  number={6},
  pages={065003},
  year={2022},
  publisher={APS}
}

@article{michel2014dynamic,
  title={Dynamic control of the polarization of intense laser beams via optical wave mixing in plasmas},
  author={Michel, Pierre and Divol, Laurent and Turnbull, David and Moody, John D},
  journal={Physical Review Letters},
  volume={113},
  number={20},
  pages={205001},
  year={2014},
  publisher={APS}
}

@article{edwards2016waveform,
  title={Waveform-controlled relativistic high-order-harmonic generation},
  author={Edwards, Matthew R and Mikhailova, Julia M},
  journal={Physical Review Letters},
  volume={117},
  number={125001},
  pages={10},
  year={2016}
}

@inproceedings{fasano2024cascaded,
  title={Cascaded Plasma Mirrors for Two-Color-Driven Harmonic Generation},
  author={Fasano, Nicholas M and Dewan, Vedin and Edwards, Matthew R and Giakas, Andreas and Bennett, Timothy and Mikhailova, Julia M},
  booktitle={2024 Conference on Lasers and Electro-Optics (CLEO)},
  pages={1--2},
  year={2024},
  organization={IEEE}
}

@inproceedings{fasano2024plasma,
  title={Plasma Mirrors for Generating Co-and Counter-Rotating Harmonics},
  author={Fasano, NM and Dewan, V and Mikhailova, JM},
  booktitle={2024 Conference on Lasers and Electro-Optics (CLEO)},
  pages={1--2},
  year={2024},
  organization={IEEE}
}

@article{edwards2023control,
  title={Control of intense light with avalanche-ionization plasma gratings},
  author={Edwards, M. R. and Waczynski, Stefan and Rockafellow, Ela and Manzo, Lili and Zingale, Anthony and Michel, Pierre and Milchberg, H. M.},
  journal={Optica},
  volume={10},
  number={12},
  pages={1587--1594},
  year={2023},
  publisher={Optica Publishing Group}
}

@article{durand2012dynamics,
  title={Dynamics of plasma gratings in atomic and molecular gases},
  author={Durand, Magali and Jarnac, Am{\'e}lie and Liu, Yi and Prade, Bernard and Houard, Aur{\'e}lien and Tikhonchuk, Vladimir and Mysyrowicz, Andr{\'e}},
  journal={Physical Review E},
  volume={86},
  number={3},
  pages={036405},
  year={2012},
  publisher={APS}
}

@article{zhang2021ionization,
  title={Ionization induced plasma grating and its applications in strong-field ionization measurements},
  author={Zhang, Chaojie and Nie, Zan and Wu, Yipeng and Sinclair, Mitchell and Huang, Chen-Kang and Marsh, Ken A and Joshi, Chan},
  journal={Plasma Physics and Controlled Fusion},
  volume={63},
  number={9},
  pages={095011},
  year={2021},
  publisher={IOP Publishing}
}

@article{moharam1978criterion,
  title={{Criterion for Bragg and Raman-Nath diffraction regimes}},
  author={Moharam, M Gamal and Young, L},
  journal={Applied Optics},
  volume={17},
  number={11},
  pages={1757--1759},
  year={1978},
  publisher={Optical Society of America}
}

@article{sheng2003plasma,
  title={Plasma density gratings induced by intersecting laser pulses in underdense plasmas},
  author={Sheng, Z-M and Zhang, Jiandi and Umstadter, Donald},
  journal={Applied Physics B},
  volume={77},
  pages={673--680},
  year={2003},
  publisher={Springer}
}

@article{lehmann2016transient,
  title={Transient plasma photonic crystals for high-power lasers},
  author={Lehmann, Goetz and Spatschek, Karl-Heinz},
  journal={Physical Review Letters},
  volume={116},
  number={22},
  pages={225002},
  year={2016},
  publisher={APS}
}

@article{shi2011generation,
  title={{Generation of high-density electrons based on plasma grating induced Bragg diffraction in air}},
  author={Shi, Liping and Li, Wenxue and Wang, Yongdong and Lu, Xin and Ding, Liang’en and Zeng, Heping},
  journal={Physical Review Letters},
  volume={107},
  number={9},
  pages={095004},
  year={2011},
  publisher={APS}
}

@article{suntsov2009femtosecond,
  title={Femtosecond laser induced plasma diffraction gratings in air as photonic devices for high intensity laser applications},
  author={Suntsov, S and Abdollahpour, D and Papazoglou, DG and Tzortzakis, S},
  journal={Applied Physics Letters},
  volume={94},
  number={25},
  year={2009},
  publisher={AIP Publishing}
}

@misc{yeh1994introduction,
  title={Introduction to photorefractive nonlinear optics},
  author={Yeh, Pochi and Moerner, WE},
  year={1994},
  publisher={American Institute of Physics}
}

@article{Edwards2020x,
  author  = {M. R. Edwards and J. M. Mikhailova},
  journal = {Scientific Reports},
  title   = {The X-Ray Emission Effectiveness of Plasma Mirrors: Reexamining Power-Law Scaling for Relativistic High-Order Harmonic Generation},
  year    = {2020},
  pages   = {5154},
  volume  = {10},
}

@article{takeda1982fourier,
  title={Fourier-transform method of fringe-pattern analysis for computer-based topography and interferometry},
  author={Takeda, Mitsuo and Ina, Hideki and Kobayashi, Seiji},
  journal={Journal of the Optical Society of America},
  volume={72},
  number={1},
  pages={156--160},
  year={1982},
  publisher={Optical Society of America}
}

@article{liu2011two,
  title={Two-dimensional plasma grating by non-collinear femtosecond filament interaction in air},
  author={Liu, Jia and Li, Wenxue and Pan, Haifeng and Zeng, Heping},
  journal={Applied Physics Letters},
  volume={99},
  number={15},
  year={2011},
  publisher={AIP Publishing}
}

@article{lemos2018guiding,
  title={Guiding of laser pulses in plasma waveguides created by linearly-polarized femtosecond laser pulses},
  author={Lemos, N and Cardoso, L and Geada, J and Figueira, G and Albert, F and Dias, JM},
  journal={Scientific Reports},
  volume={8},
  number={1},
  pages={3165},
  year={2018},
  publisher={Nature Publishing Group UK London}
}

@article{wu2005manipulating,
  title={{Manipulating ultrashort intense laser pulses by plasma Bragg gratings}},
  author={Wu, Hui-Chun and Sheng, Zheng-Ming and Zhang, Qiu-Ju and Cang, Yu and Zhang, Jie},
  journal={Physics of Plasmas},
  volume={12},
  number={11},
  year={2005},
  publisher={AIP Publishing}
}

@article{lehmann2024plasma,
  title={Plasma-grating-based laser pulse compressor},
  author={Lehmann, G and Spatschek, KH},
  journal={Physical Review E},
  volume={110},
  number={1},
  pages={015209},
  year={2024},
  publisher={APS}
}

@article{jarnac2014study,
  title={Study of laser induced plasma grating dynamics in gases},
  author={Jarnac, Am{\'e}lie and Durand, Magali and Liu, Yi and Prade, Bernard and Houard, Aur{\'e}lien and Tikhonchuk, Vladimir and Mysyrowicz, Andr{\'e}},
  journal={Optics Communications},
  volume={312},
  pages={35--42},
  year={2014},
  publisher={Elsevier}
}

@article{palastro2015plasma,
  title={Plasma lenses for ultrashort multi-petawatt laser pulses},
  author={Palastro, JP and Gordon, D and Hafizi, B and Johnson, LA and Pe{\~n}ano, J and Hubbard, RF and Helle, M and Kaganovich, D},
  journal={Physics of Plasmas},
  volume={22},
  number={12},
  year={2015},
  publisher={AIP Publishing}
}

@article{qu2017plasma,
  title={Plasma q-plate for generation and manipulation of intense optical vortices},
  author={Qu, Kenan and Jia, Qing and Fisch, Nathaniel J},
  journal={Physical Review E},
  volume={96},
  number={5},
  pages={053207},
  year={2017},
  publisher={APS}
}

@article{maksimchuk2025zeus,
  title={The \text{ZEUS} multi-petawatt laser system},
  author={Maksimchuk, A and Nees, J and Hou, B and Anthony, R and Bae, J and Bayer, F and Burger, M and Campbell, PT and Cardarelli, J and Contreras, V and others},
  journal={Physics of Plasmas},
  volume={32},
  number={10},
  year={2025},
  publisher={AIP Publishing}
}

@article{polyanskiy2020demonstration,
  title={{Demonstration of a 2 ps, 5 TW peak power, long-wave infrared laser based on chirped-pulse amplification with mixed-isotope CO$_{2}$ amplifiers}},
  author={Polyanskiy, Mikhail N and Pogorelsky, Igor V and Babzien, Marcus and Palmer, Mark A},
  journal={OSA Continuum},
  volume={3},
  number={3},
  pages={459--472},
  year={2020},
  publisher={Optical Society of America}
}

@article{thaury2007plasma,
  title={Plasma mirrors for ultrahigh-intensity optics},
  author={Thaury, C{\'e}dric and Quere, Fabien and Geindre, J-P and Levy, Anna and Ceccotti, Tiberio and Monot, P and Bougeard, Michel and R{\'e}au, F and d’Oliveira, P and Audebert, Patrick and others},
  journal={Nature Physics},
  volume={3},
  number={6},
  pages={424--429},
  year={2007},
  publisher={Nature Publishing Group UK London}
}

@article{branlard2012european,
  title={{The european XFEL LL RF system}},
  author={Branlard, Julien and Ayvazyan, Gohar and Ayvazyan, Valeri and Grecki, Mariusz and Hoffmann, M and Jezynski, T and Kudla, IM and Lamb, T and Ludwig, F and Mavric, U and others},
  journal={IPAC},
  volume={12},
  pages={55--57},
  year={2012}
}

@article{edwards2021laser,
  title={Laser-driven plasma sources of intense, ultrafast, and coherent radiation},
  author={Edwards, Matthew R and Fisch, Nathaniel J and Mikhailova, Julia M},
  journal={Physics of Plasmas},
  volume={28},
  number={1},
  year={2021},
  publisher={AIP Publishing}
}

@inproceedings{wang2025experimental,
  title={Experimental Demonstration of Chromatic Angular Dispersion from Transmission Plasma Gratings},
  author={Wang, Michelle M and Perez-Ramirez, Victor M. and Das, Arunava and Tigges-Green, I and Dewan, V and Ou, K and Cao, S and Michel, P and Edwards, Matthew R and Mikhailova, Julia M},
  booktitle={2025 Conference on Lasers and Electro-Optics (CLEO)},
  pages={1--2},
  year={2025},
  organization={IEEE}
}
\end{document}